\documentclass[twocolumn,showpacs,aps,pra,letter]{revtex4-1}
\usepackage{amssymb}
\usepackage{amsfonts}
\usepackage{dcolumn}
\usepackage{amsmath}
\usepackage{graphicx}
\usepackage{psfrag}
\usepackage{amsmath}
\usepackage{amssymb}
\usepackage{epstopdf}
\newcommand{\be} {\begin{equation}}
\newcommand{\ee} {\end{equation}}

\begin{document}

\title{High-efficiency quantum state transfer and quantum memory using a mechanical oscillator}
\author{Eyob A. Sete$^1$ and H. Eleuch$^{2,3}$}
\affiliation{$^1$Department of Electrical Engineering, University of California, Riverside, California 92521, USA\\
$^2$Department of Physics, McGill University, Montreal, Canada H3A 2T8\\
$^{3}$Department of Physics, Universit\'{e} de Montr\'{e}al, Montreal, QC,  H3T 1J4, Canada}
\date{\today}

\begin{abstract}
We analyze an optomechanical system that can be used to efficiently transfer a quantum state between an optical cavity and a distant mechanical oscillator coupled to a second optical cavity. We show that for a moderate mechanical Q-factor it is possible to achieve a transfer efficiency of $99.4\%$ by using adjustable cavity damping rates and destructive interference. We also show that the quantum mechanical oscillator can be used as a quantum memory device with an efficiency of $96\%$ employing a pulsed optomechanical coupling. Although the mechanical dissipation slightly decreases the efficiency, its effect can be significantly reduced by designing a high-Q mechanical oscillator.
\end{abstract}
\pacs{42.50.Ex, 07.10.Cm, 42.50.Wk}
\maketitle
\section{Introduction}
 Realization of a high fidelity quantum state transfer between two nodes of a quantum network is a crucial task in quantum information processing \cite{Kim08,DiVincenzo-00}. Hybrid quantum systems involving atoms \cite{Cir97,Razavi-06}, ions \cite{Rem12}, superconducting circuits \cite{Jah07,Kor11,Sri14,Wen14,Set14}, and quantum mechanical oscillators \cite{Cle12,Tia12,Mcg13,Pal13} are shown to be useful in realizing a high fidelity transfer. In particular, hybrid systems consisting of electromagnetic and mechanical oscillators are promising platforms for transferring a quantum state from site to site or for a quantum memory. Quantum state transfer between optical and microwave cavities mediated by a quantum mechanical oscillator has been proposed \cite{Cle12,Tia12,Mcg13,Pal13}. Nevertheless, in all of these studies the two electromagnetic cavities are not physically far apart. Implementation of quantum state transfer between remote resonators or qubits using flying qubits in a quantum network is desirable and opens the possibility for a variety of novel applications such as entangled-state cryptography \cite{Ben91}, teleportation \cite{Ben93}, and purification \cite{Ben96} among others.  Besides, as being a bridge to interface between two electromagnetic cavities, quantum mechanical oscillators can be used as a quantum memory device and a frequency transducer \cite{Mcg13}.

 In this work, we propose a scheme for the transfer of a quantum state between an optical cavity and a distant mechanical oscillator coupled to an optical cavity via radiation pressure. We refer to this procedure as a ``writing'' protocol. The transfer is mediated by a flying qubit and can be realized with an efficiency arbitrarily close to unity. We also analyze the reverse process in which a quantum state stored in the mechanical mode transferred to the optical mode in a distant cavity; we refer to this procedure as a ``reading'' protocol. Such a high efficiency state transfer is achieved by using time-varying cavity damping rates and destructive interference. Modulation of the cavity damping rate can also be used to speed up the cooling process while suppressing the heating noise \cite{Liu13}. Using experimental parameters \cite{Gro09} and for a moderate mechanical quality factor $Q=6,700$, a transfer efficiency as high as $99.4\%$ can be achieved. We also propose a quantum memory device using the mechanical oscillator's degrees of freedom. We show that due to the long life time (in ms time scale) of the mechanical mode, it is possible to build a quantum memory with a reasonably long storage time. Rapid retrieval of the stored quantum state can be realized by designing appropriate optomechanical coupling pulses. We find that for the moderate mechanical Q-factor, a quantum state can be stored and retrieved with $96\%$ efficiency. We emphasize that even though we considered quantum state transfer between two optical cavities, the idea can be extended to the state transfer between two microwave resonators using superconducting quantum circuits or other experimental setups described in Refs. \cite{Hof09,Lov10,Kak10,Wal04}.

 \section{Model and transfer protocols}\label{model}
We consider an optical cavity spatially separated from a second optical cavity that is coupled to a mechanical oscillator. We propose and analyze two state transfer protocols. In the first protocol (Fig. \ref{fig1}) we consider a quantum state encoded onto the optical mode of the first cavity and transfer it to the mechanical mode coupled to the second cavity via radiation pressure. We refer to this process as a ``writing" protocol, because the quantum state is mapped onto the mechanical degrees of freedom. In the second protocol (see Fig. \ref{fig11}), a quantum state stored in the mechanical oscillator that is coupled to the second cavity mode is transferred to the optical mode in the first cavity via a flying qubit (propagating photon). We refer to this process as a ``reading'' protocol. In both protocols, to avoid multiple reflections between the two cavities, we use a Faraday's isolator so that the reflected field at the second cavity will be directly sent to a detector.

In order to reduce the dissipation during the photon propagation in the transmission channel, the two cavities can be coupled via an optical fiber (for case of optical cavities) or by a microwave superconducting transmission line (for microwave cavities). Recently, an efficient coupling between an optical fiber and a cavity has been realized \cite{Osk13,Coh13}.

A high efficiency quantum state transfer can be realized by cancelling the back-reflection at the second cavity via destructive interference \cite{Kor11,Set14}. An almost perfect cancellation can be achieved by either time-varying the damping rate of the first cavity, second cavity, or both. In the following we discuss the ``writing'' and ``reading'' protocols by using time-varying cavity damping rates.

\subsection{The ``writing'' protocol: quantum state transfer from an optical cavity to a mechanical oscillator}
\begin{figure}
\includegraphics[width=8.5cm]{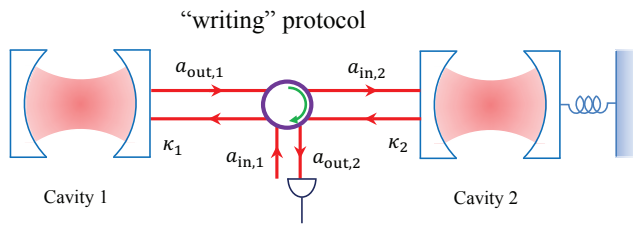}
\caption{Schematic of the ``writing'' protocol. A quantum state initially encoded on the cavity mode 1 is transferred to the mechanical mode coupled to cavity 2 via radiation pressure. Nearly perfect state transfer can be achieved by using adjustable cavity  damping rates $\kappa_{1}$ and $\kappa_{2}$. The output field $a_{1,\rm out}$ from the first cavity propagates away and becomes an input to the second cavity. To avoid multiple refections between the cavities, we use a Faraday's isolator. This allows the output $a_{2,\rm out}$ of the second cavity to be directly go to the detector. }\label{fig1}
\end{figure}

Here we consider an initial quantum state encoded on the optical mode of cavity 1 and transfer it to the mechanical mode in the second cavity via a flying qubit (see Fig. \ref{fig1}). This can be interpreted as ``writing'' a quantum state onto the mechanical degrees of freedom, which can later be retrieved using the opposite procedure---``reading'' (transfer of a quantum state from mechanical to optical mode).

In the interaction picture and in a displaced frame with respect to the classical mean value in each cavity, the Hamiltonian can be written as \cite{Set10,Set11,Set12,Set13}
\begin{align}\label{NH}
  H=\omega_{\rm M}b^{\dag}b-\Delta_{1} a_{1}^{\dag}a_{1}-\Delta_{2} a_{2}^{\dag}a_{2}+G(a_{2}b^{\dag}+a^{\dag}_{2}b),
  \end{align}
where $\omega_{\rm M}$ is the mechanical frequency, $b$ is the annihilation operator for the mechanical mode, $G$ is the many-photon optomechanical coupling, $\Delta_{j}$ is the detuning of the optical drive applied to the $j$th cavity, $a_{j}$ is the annihilation operator for the $j$th cavity mode.
Using this Hamiltonian, we obtain the quantum Langevin equations for the cavity and mechanical mode operators
\begin{align}\label{ql1}
  \dot a_{1}&=-\frac{\kappa_{1}}{2}a_{1}+i\Delta_{1} a_{1}+\sqrt{\kappa_{1}} a_{\rm in,1},\\
\label{ql2}
 \dot b&=-\frac{\gamma}{2}b-i\omega_{\text{M}}b-iGa_{2}+\sqrt{\gamma}f,\\
\label{ql3}
  \dot a_{2}&=-\frac{\kappa_{2}}{2}a_{2}+i\Delta_{2} a_{2}-iGb+\sqrt{\kappa_{2}} a_{\rm in, 2},
\end{align}
where $\gamma$  is the mechanical oscillator damping rate, $\kappa_{j}$ is the $j$th cavity damping rate, $f$ is the noise operator for the mechanical oscillator bath, while $a_{\rm in, j}$ describes the quantum noise of the vacuum field incident on the $j$th cavity. Note that due to the unidirectional coupling the output of the cavity 1 is an input to cavity 2 with appropriate time delay, that is, $a_{\rm out, 1}(t-\tau)\equiv a_{\rm in, 2}(t)$. Thus, using input-output relation for cavity 1: $a_{\rm out, 1}(t)=\sqrt{\kappa_{1}}a_{1}(t)-a_{\rm in,1}(t)$, the input noise operator at cavity 2 is given by
\begin{equation}
a_{\rm in, 2}(t)=\sqrt{\kappa_{1}}a_{1}(t-\tau)-a_{\rm in, 1}(t-\tau).
\end{equation}
In view of this, the equation for the cavity mode operator $a_{2}$ becomes
\begin{align}\label{ql33}
  \dot a_{2}=&-\frac{\kappa_{2}}{2}a_{2}-i\Delta_{2}a_{2}-iGb +\sqrt{\kappa_{1}\kappa_{2}} a_{1}(t-\tau)\notag\\
  &-\sqrt{\kappa_{2}} a_{\rm in, 1}(t-\tau).
\end{align}

It is worth noting that due to unidirectional coupling Eqs. \eqref{ql1} is decoupled from \eqref{ql33}. This unidirectional coupling is an example of the so-called quantum cascade system \cite{Gar93,Car93}. The time delay $\tau$ can be eliminated by defining a ``time-delayed'' operators for the first cavity and for the mechanical oscillator, for example, $\tilde a_{1}(t)\equiv a_{1}(t-\tau)$, $\tilde b(t)\equiv b(t-\tau)$. In the following we assume that we have performed these transformations and for simplicity we drop the tilde, which amounts to $\tau\rightarrow 0$ in all equations.

Note that, in principle, the detunings $\Delta_{j}$ can be arbitrarily chosen provided that the system remains stable. However, since we are interested in quantum state transfer from one system to the other (in our case from cavity 1 to the mechanical oscillator) it is necessary to choose cavity-laser detuning to be tuned at the mechanical frequency, i.e., $\Delta_{j}=\omega_{\rm M}$. Thus, in a frame rotating with $\omega_{\rm M}$ and choosing $\Delta_{j}=\omega_{\rm M}$,  Eqs. \eqref{ql1}, \eqref{ql2}, and \eqref{ql33} reduce to

\begin{eqnarray}
\label{g11}
 \dot a_{1}&=&-\frac{\kappa_{1}}{2} a_{1}+\sqrt{\kappa_{1}}a_{\rm in,1},\\
\label{g12}
  \dot b&=&-\frac{\gamma}{2} b-iG a_{2}+\sqrt{\gamma}f,\\
  \label{g13}
  \dot a_{2}&=&-\frac{\kappa_{2}}{2} a_{2}-iG b +\sqrt{\kappa_{1}\kappa_{2}} a_{1}-\sqrt{\kappa_{2}}a_{\rm in,1}.
  \end{eqnarray}

It has been shown that \cite{Jah07,Kor11,Set14} a high efficiency quantum state transfer between two remote cavities connected by a transmission line can be realized by cancelling the back reflection at the receiving cavity. This can be achieved by designing time-varying cavity 1, cavity 2, or both damping rates so that the field emitted has an exponentially increasing/decreasing waveform that allows destructive interference at the receiving cavity. In this paper, following Ref. \cite{Kor11}, we divide the protocol in two parts. In the first part, we assume the first cavity damping rate $\kappa_{1}$ varies in time, while the damping rate of the second cavity is fixed at its maximum value, $\kappa_{2,\rm m}$. In order to derive the time-profile of the damping rate $\kappa_{1}$, we assume a simpler problem: state transfer between two cavities. Thus the formal solution of Eq. \eqref{g13}, after dropping the $G$ term and assuming the cavities are coupled to a vacuum environment, has the form
\begin{align}\label{fs}
a_{2}(t)&=a_{2}(0)e^{-\kappa_{2,\rm m} t/2}+a_{1}(0)\sqrt{\kappa_{2,\rm m}}\notag\\
&\times \int_{0}^{t}dt'\sqrt{\kappa_{1}(t')}e^{-\kappa_{2,\rm m}(t-t')/2} e^{-\frac{1}{2}\int_{0}^{t'}\kappa_{1}(t'')dt''}.
\end{align}
If the state is initially encoded onto the mode of the first cavity $a_{1}(0)$ and the second cavity is initially in vacuum state, the state transfer efficiency can be defined as \cite{Jah07}
\begin{align}
\sqrt{\eta_{1}}=\sqrt{\kappa_{2,\rm m}}\int_{0}^{t}dt'\sqrt{\kappa_{1}(t')}e^{-\kappa_{2,\rm m}(t-t')/2} e^{-\frac{1}{2}\int_{0}^{t'}\kappa_{1}(t'')dt''}.
\end{align}
We next search for optimum pulse shapes of the damping rate $\kappa_{1}$ that maximizes the efficiency $\eta_{1}$ for the first part of the procedure. This can be done using the Euler-Lagrange formalism. The damping rates satisfy the differential equation \cite{Jah07}
\begin{equation}\label{kap1}
\kappa _{1}^{2}(t)-\dot{\kappa }_{1}(t)+\kappa_{2,\rm m}\kappa _{1}(t)=0.
\end{equation}
Therefore, during the first part of the procedure $0\leq t\leq t_{\rm m}$, the damping rat $\kappa_{1}(t)$ obtained by solving Eq. \eqref{kap1} is given by
\begin{align}
\kappa _{1}(t)&=\frac{\kappa_{1, \rm m}}{2e^{\kappa_{2,\rm m}(t_{\rm m}-t)}-1},\label{k1}
\end{align}
where $\kappa_{1,\rm m}$ is the maximum value of the first cavity damping rate.

In the second part of the procedure, the damping rate of the first cavity remain constat, $\kappa _{1}=\kappa_{1,\rm m}$, while the damping rate of the receiving cavity decreases in time. Using the time reversal symmetry, the pulse shape of the damping rate $\kappa_{2}$ which maximizes the efficiency $\eta_{2}$ of the second part of the procedure $t\geq t_{\rm m}$ can be written as
\begin{align}\label{k2}
\kappa_{2}(t)=\frac{\kappa_{2,\rm m}}{2e^{\kappa_{1,\rm m}(t-t_{\rm m})}-1}.
\end{align}
The derivation of the pulse shapes \eqref{k1} and \eqref{k2} of the damping rates assume the state transfer between two cavities. In the following, we use the same pulse shapes to analyze the transfer of a quantum state between the first cavity to the mechanical oscillator coupled to the second cavity mode. This is because the state has to first be transferred before it is mapped to the mechanical oscillator.

Note that the pulse shapes Eqs. \eqref{k1} and \eqref{k2} were also derived invoking the destructive interference condition for the cancellation of the back-reflected field into the environment \cite{Kor11,Set14}.
To analyze the quantum state transfer when the cavities and the mechanical oscillators are coupled to vacuum (zero temperature environment), it is sufficient to use the corresponding classical equations for Eqs. \eqref{g11}-\eqref{g13}
\begin{eqnarray}
\label{l11}
 \dot \alpha_{1}&=&-\frac{\kappa_{1}(t)}{2} \alpha_{1},\\
\label{l12}
  \dot \beta &=&-\frac{\gamma}{2} \beta-iG \alpha_{2},\\
  \label{l13}
  \dot \alpha_{2}&=&-\frac{\kappa_{2}(t)}{2} \alpha_{2}-iG \beta +\sqrt{\kappa_{1}(t)\kappa_{2}(t)} \alpha_{1}.
  \end{eqnarray}
These equations will be used to analyze the quantum state transfer from cavity 1 to the mechanical oscillator.

\begin{figure}[t]
(a)~~%\\
\includegraphics[width=8cm]{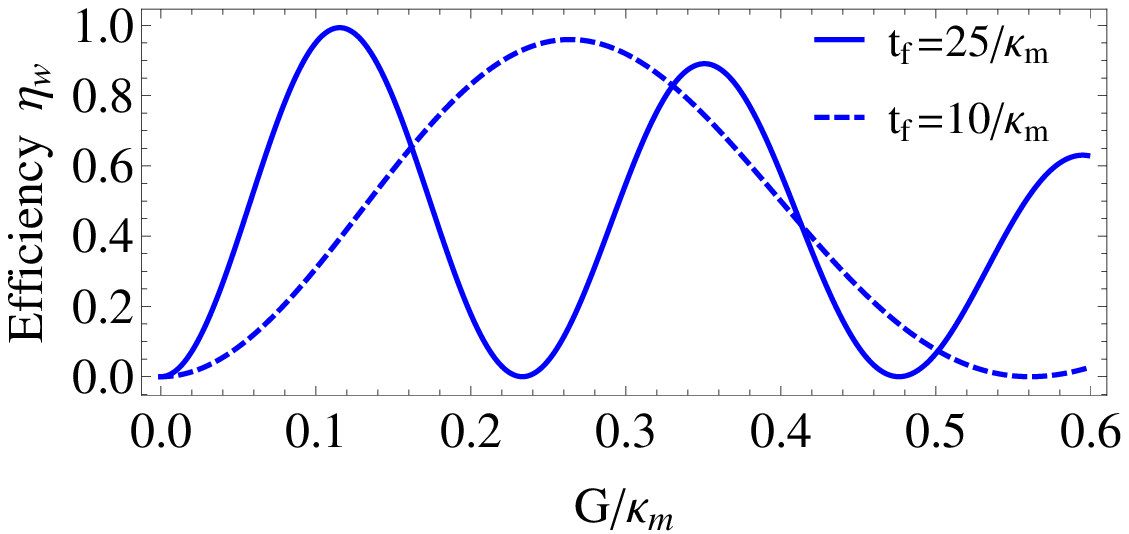}
(b)~~%\\
\includegraphics[width=8cm]{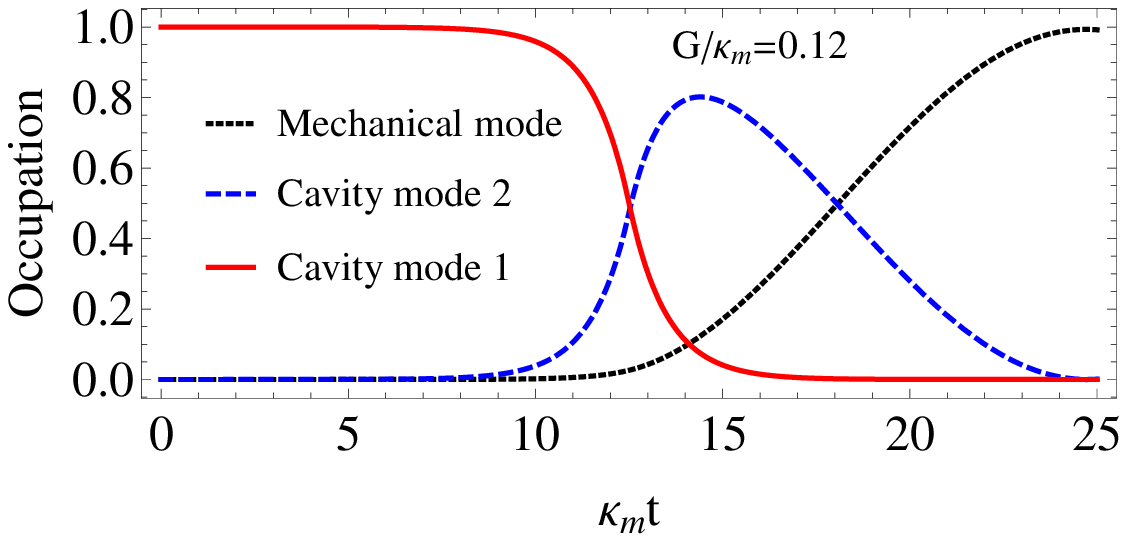}
(c)~~%\\
\includegraphics[width=8cm]{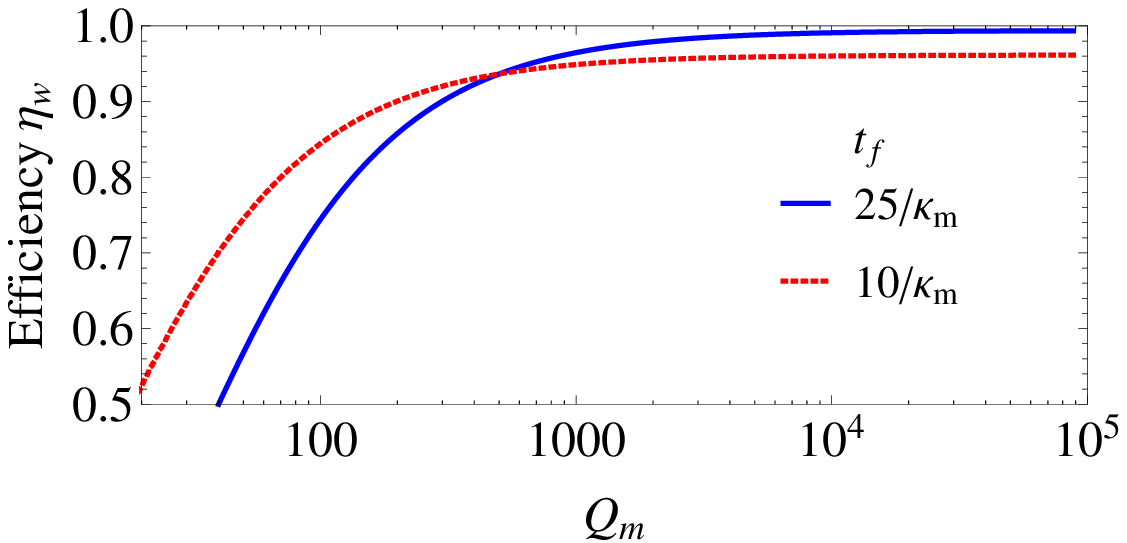}
\caption{(a)The efficiency $\eta_{\rm w}$ of the state transfer from the first cavity mode to the mechanical mode optomechanically coupled to the second cavity mode as a function of $G/\kappa_{\rm m}$ for $\gamma/\kappa_{\rm m}=6.5\times 10^{-4}$ and for the procedure times $t_{\rm f}=10/\kappa_{\rm m}$ (dashed curve) and $25/\kappa_{\rm m}$ (solid curve). (b) Plot of the occupation number for the three modes as a function of the scaled time $\kappa_{\rm m}t$ for the optomechanical coupling where the efficiency is maximum $G/\kappa_{\rm m}=0.12$ and for the procedure time $t_{\rm f}=25/\kappa_{\rm m}$. (c) Efficiency versus the mechanical quality factor $Q_{\rm m}=\omega_{\rm M}/\gamma$ for procedure time and the optomechanical coupling $(t_{\rm f}, G)=(25/\kappa_{\rm m},0.12/\kappa_{\rm m})$ (blue solid curve) and $(t_{\rm f}, G)=(10/\kappa_{\rm m},0.265/\kappa_{\rm m})$ (red dashed curve). See the text for other parameters. }\label{fig2}
\end{figure}

We numerically solve the above equations, assuming the initial condition $\alpha_{1}(0)=1$, $\beta(0)=0$, and $\alpha_{2}(0)=0$ and characterize the performance of the state transfer from the first cavity mode to the mechanical mode by the energy transfer efficiency $\eta_{\rm w}=|\beta(t_{\rm f})|^2/|\alpha_1(0)|^2$. Unless mentioned otherwise, we assume $\kappa_{1,\rm m}=\kappa_{2,\rm m}\equiv\kappa_{\rm m}$. In the numerical integration we assume two identical cavities and use the parameters from recent optomechanical experiment \cite{Gro09}: laser frequency $\omega_{L}=2\pi\times 2.82\times 10^{14}~\text{Hz} ~(\lambda=1064 ~\text{nm})$, cavity frequency $\omega_{r}=2\pi\times 5.64\times 10^{14}~\text{Hz}~(\omega_{r}=2\omega_{L})$, maximum cavity damping rate $\kappa_{\rm m}=2\pi\times 215~ \text{kHz}$, mechanical damping rate $\gamma=2\pi\times 140~\text{Hz}$, mechanical frequency $\omega_{\rm M}=2\pi\times 947 ~\text{kHz}$. In Fig. \ref{fig2} a, we show the state transfer efficiency as a function of the scaled many-photon optomechanical coupling, $G/\kappa_{\rm m}$ for different values of procedure time $t_{\rm f}$. For the above parameters and for a procedure time of $t_{\rm f}=25/\kappa_{\rm m}$ the transfer efficiency can be as high as $99.4\%$ at $G=0.12\kappa_{\rm m}$. We see that the obtained efficiency exhibit damped oscillations, decreasing with increasing the many-photon optomechanical coupling, which is somewhat counter intuitive. Moreover, the transfer efficiency increases with increasing procedure time. This is because the inefficiency of the procedure depends not only on the initial loss during the buildup time of the field in the second cavity, but also on the untransmitted field that remains in first cavity when the protocol is abruptly stopped at $t_{\rm f}$. Thus the longer the procedure time, the less the amount of energy left in first cavity and better the efficiency. It was shown that the inefficiency of the transfer protocol is related to the procedure time by $1-\eta_{\rm w}\approx \exp(-\kappa_{\rm m}t_{\rm f}/2)$ \cite{Set14}.

The evolution of the occupation number for each mode is shown in Fig. \ref{fig2} (b) for procedure time $t_{\rm f}=25/\kappa_{\rm m}$ and choosing the optomechanical coupling that gives the maximum transfer, $G/\kappa_{\rm m}=0.12$. As can be seen the initial state in cavity 1 is first transferred mostly to the second cavity mode and then slowly mapped to the mechanical mode with an efficiency $99.4\%$. The transfer efficiency strongly depends on the mechanical quality factor $Q_m=\omega_{\rm M}/\gamma$; see Fig. \ref{fig2} (c). The efficiency increases with the mechanical quality factor and saturates at $99.4\%$ for $Q_{m}\gtrsim 7\times 10^3$. The transfer efficiency is higher for longer procedure times and for large values of the mechanical Q-factor.

So far we assumed that the transmission channel between the two cavities is lossless and ignored the intrinsic dissipations of the cavities, $1/T_{1, \rm cav 1}$ and $1/T_{1, \rm cav 2}$. In general, the transmission channel suffers from different losses such as cavity-fiber coupling loss and intrinsic fiber optical loss. To estimate these losses we introduce the energy efficiency of the transmission channel $\eta_{\rm tr}$. As a result, Eqs. \eqref{l11} and \eqref{l13} now read $\dot \alpha_{1}=-(\kappa_{1}(t)+1/T_{1, \rm cav 1})\alpha_{1}/2$,
$\dot\alpha_{2}=-(\kappa_{2}(t)+1/T_{1, \rm cav 2})\alpha_{2}/2-iG\beta+\sqrt{\kappa_{1}(t)\kappa_{2}(t)}\sqrt{\eta_{\rm tr}}\alpha_{1}$. We analyze the contribution of each loss separately. The transmission channel loss yields a contribution to the transfer efficiency as $\eta_{\rm w}\rightarrow \eta_{\rm w}\eta_{\rm tr}$. For $1/T_{1, \rm cav 1},1/T_{1, \rm cav 2}\ll k_{\rm m}$, the contribution of the dissipations decrease the transfer efficiency exponentially, $\eta_{\rm w}\rightarrow \eta_{\rm w}\exp(-t_{\rm f}/2T_{1, \rm cav 1})\exp(-t_{\rm f}/2T_{1, \rm cav 2})$. These estimates apply to the ``reading'' and the quantum memory protocol discussed below.

\subsection{The ``reading'' protocol: quantum state transfer from a mechanical oscillator to an optical cavity}
\begin{figure}[t]
\includegraphics[width=8cm]{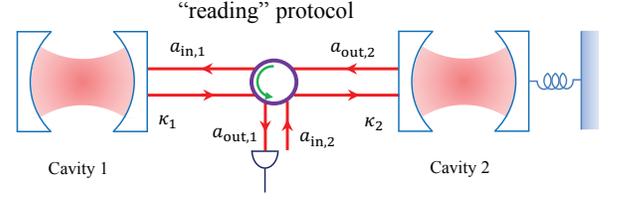}
\caption{Schematic of the ``reading'' protocol. A quantum state initially stored in the mechanical oscillator is transferred to cavity 1 via flying qubit. Nearly perfect state transfer can be achieved by using adjustable cavity damping rates $\kappa_{1}$ and $\kappa_{2}$. The time profiles of the damping rates are now interchanged compared to the writing protocol (see the text).}\label{fig11}
\end{figure}

In the previous section we discussed how to transfer a quantum state from an optical cavity to a mechanical oscillator coupled to a distant cavity. Here we discuss the opposite process in which a quantum state encoded on the mechanical mode is transferred to the mode of the second cavity and then to that of the first cavity via a flying qubit. The schematic of this procedure is shown in Fig. \ref{fig11}. Notice that the output of the second cavity is now an input to the first cavity and the coupling is unidirectional. The final quantum state in the first cavity is inferred by measuring the output filed $a_{1,\rm out}$ via the method of homodyne. Following the same line of reasoning as in the ``writing'' protocol, the equations for the classical field amplitudes become
\begin{align}
\dot\alpha_{1}&=-\frac{\kappa_{1}(t)}{2}\alpha_{1}+\sqrt{\kappa_{1}(t)\kappa_{2}(t)}\alpha_{2},\label{R1}\\
\dot\beta &=-\frac{\gamma}{2}\beta-iG\alpha_{2},\\
\dot\alpha_{2} &=-\frac{\kappa_{2}(t)}{2}\alpha_{2}-i G\beta \label{R3}.
\end{align}
Note that in the reading protocol the time profiles of $\kappa_{1}$ and $\kappa_{2}$ are interchanged: $\kappa_{1}$ is maximum for $t\leq t_{\rm m}$ and decreases with time while $\kappa_{2}$ slowly increases with time and reaches it maximum value at $t=t_{\rm m}$. To characterize the quantum state transfer, we numerically solve Eqs. \eqref{R1}-\eqref{R3} and calculate the energy transfer efficiency $\eta_{\rm r}=|\alpha_{1}(t_{\rm f})|^2/|\beta(0)|^2$. In the numerical simulation, we use the initial condition $\alpha_{i}(0)=0$ and $\beta(0)=1$.

\begin{figure}[t]
(a)~~%\\
\includegraphics[width=8cm]{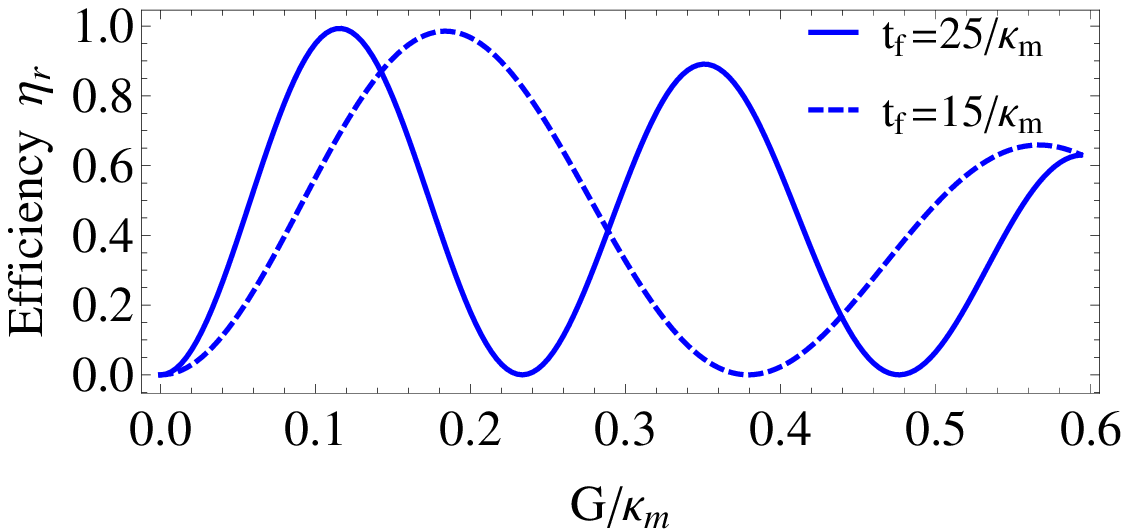}
(b)~~%\\
\includegraphics[width=8cm]{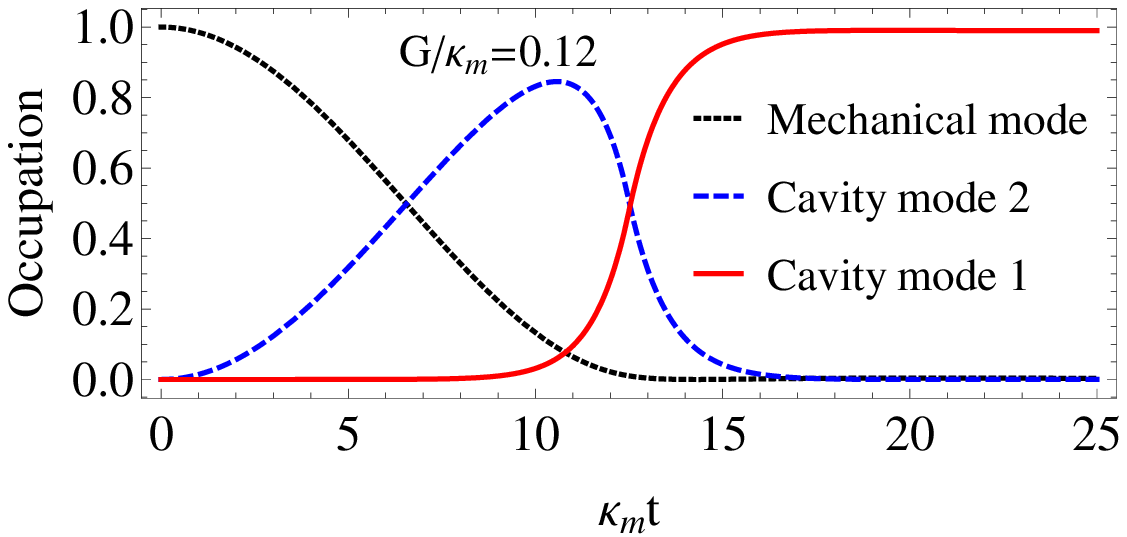}
(c)~~%\\
\includegraphics[width=8cm]{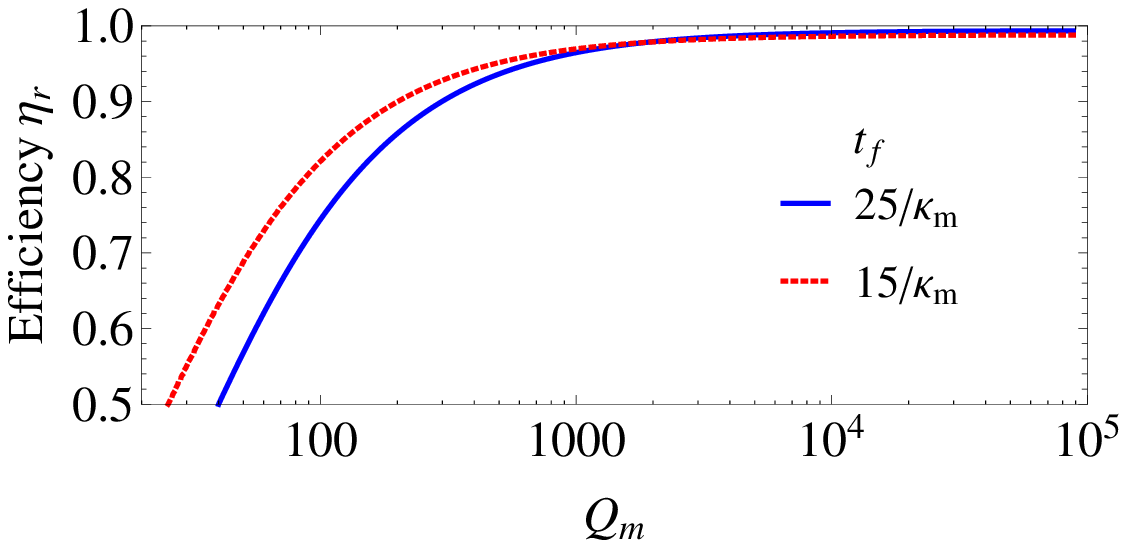}
\caption{(a) The efficiency $\eta_{\rm r}$ of the quantum state transfer from the mechanical mode to the first cavity mode as a function of the scaled optomechanical coupling $G/\kappa_{\rm m}$ and for the procedure times $t_{\rm f}=25/\kappa_{\rm m}$ (solid curve) and $15/\kappa_{\rm m}$ (dashed curve).
(b) Plot of the occupation number for the three modes as a function of the scaled time $\kappa_{\rm m}t$ for the optomechanical coupling where the efficiency is maximum $G/\kappa_{\rm m}=0.12$ and for the procedure time $t_{\rm f}=25/\kappa_{\rm m}$. (c) Efficiency versus the mechanical quality factor $Q_{\rm m}=\omega_{\rm M}/\gamma$ for the procedure time and the optomechanical coupling $(t_{\rm f}, G)=(25/\kappa_{\rm m},0.12/\kappa_{\rm m})$ (blue solid curve) and $(t_{\rm f}, G)=(15/\kappa_{\rm m},0.1835/\kappa_{\rm m})$ (red dashed curve).}\label{fig4}
\end{figure}
Figure \ref{fig4} (a) shows the state transfer efficiency from the mechanical oscillator to the first cavity for the procedure times $t_{\rm f}=15/\kappa_{\rm m}$ and $25/\kappa_{\rm m}$. Similar to the ``writing'' protocol, the efficiency exhibit damped oscillations with period of oscillation proportional to the scaled procedure time $\kappa_{\rm m}t_{\rm f}$. The efficiency increases with increasing procedure time. For the experimental parameter mentioned earlier, the maximum efficiency for procedure time $t_{\rm f}=25/\kappa_{\rm m}$ is $99.4\%$ at $G/\kappa_{\rm m}=0.12$. The plot of the occupation number for $G/\kappa_{\rm m}=0.12$ is shown in Fig. \ref{fig4} (b). We see that the energy is first transferred from the mechanical oscillator to the second cavity mode. Then, the field propagates away to the first cavity, where the quantum state is retrieved with an efficiency of $99.4\%$. Similar to the ``writing'' protocol, the efficiency strongly relies on the mechanical quality factor [see Fig. \ref{fig4}(c)].

\section{Quantum mechanical oscillator as quantum memory device}

One of the potential applications of mechanical oscillator is a quantum memory. Here we consider a process in which a quantum state encoded onto an optical mode in the first cavity and transferred and stored in the mechanical oscillator that is coupled to the second cavity. This procedure has three steps: during the first step, the quantum state is transferred to the second cavity mode via a flying qubit. This process, as discussed in the previous section, can be realized with high efficiency by using tunable cavity damping rates and destructive interference. In the second step, the quantum state, which is now encoded onto the optical mode of the second cavity, will be transferred to the mechanical mode using a pulsed coupling between the optical and the mechanical modes, $G(t)$. It is possible to make this process fast using a short coupling pulse. The quantum state which is now stored in the mechanical degrees of freedom has storage time determined by the mechanical dissipation time, which is typically few tens of ms. In the final step, the quantum state is transferred back to the optical mode of the second cavity using the pulsed coupling. Note that there is no leakage of photons from the second cavity until this stage of the procedure due to the perfect cancellation of the transmitted field into the environment by the destructive interference. Thus, to retrieve the quantum state, one has to release and measure the field in the second cavity via homodyne detection.

\begin{figure}[t]
(a)~~%\\
\includegraphics[width=8cm]{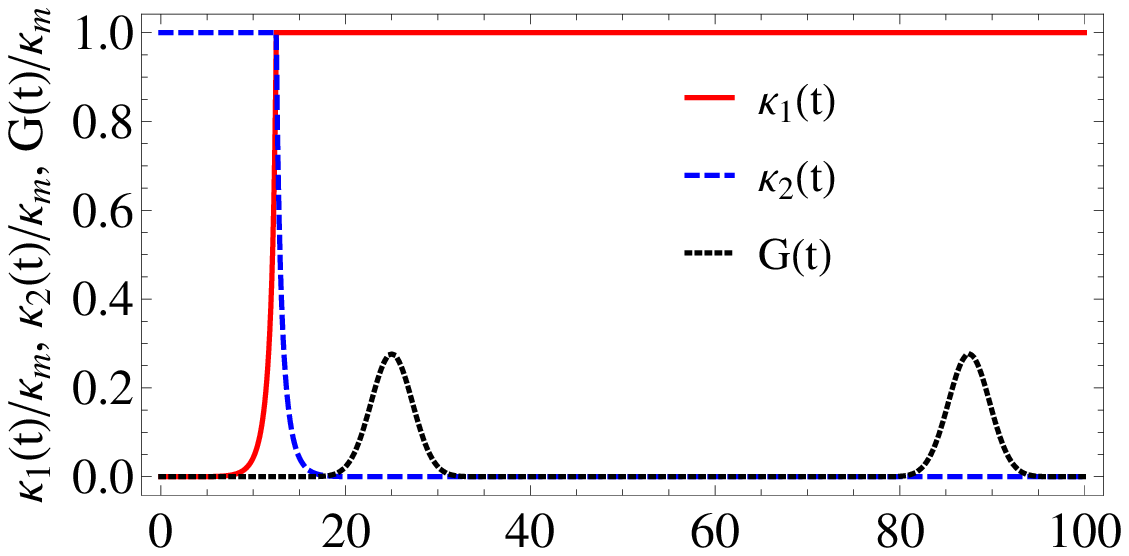}
(b)~~%\\
\includegraphics[width=8cm]{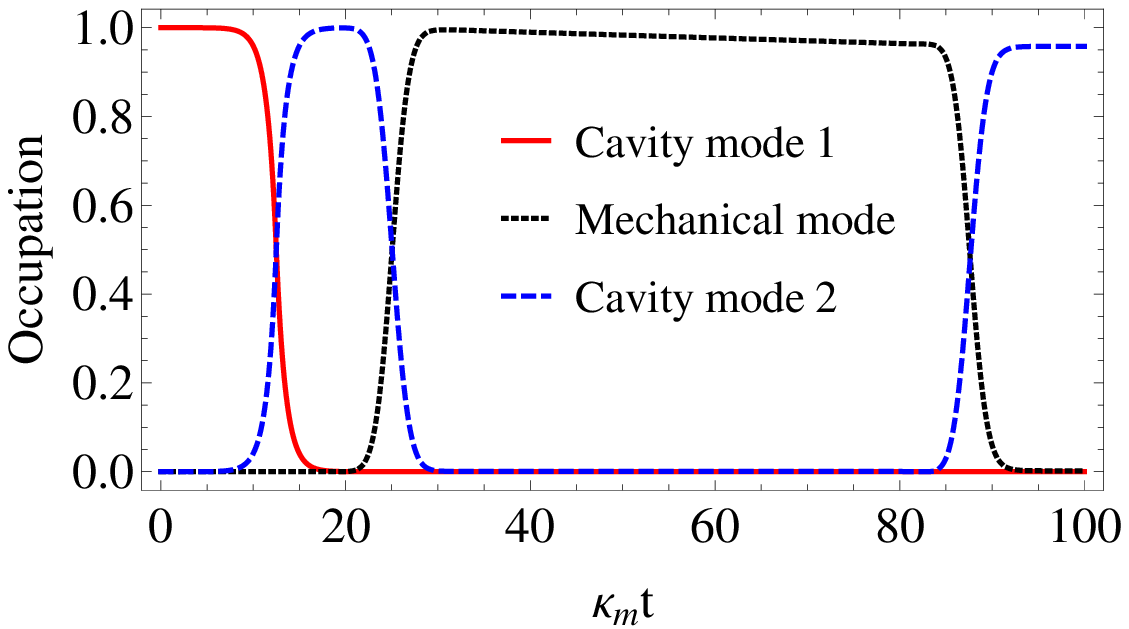}
\caption{(a) The time profiles of the first $\kappa_{1}$ (red solid curve), the second $\kappa_{2}$ (blue dashed curve) cavity damping rates normalized by the maximum value of the damping rate $\kappa_{\rm m}$, and the pulse sequence for the optomechanical coupling $G(t)$ (black dotted curve) as a function of normalized time $\kappa_{\rm m}t$. (b) Occupation number for the three modes. The quantum state is encoded on the optical mode in the first cavity (red sold curve) which is transferred via a flying qubit to the second cavity (blue dashed curve). The quantum state is then transferred to the mechanical mode (black dotted curve) using the first Gaussian pulse centered at $t_{1}=2t_{\rm m}$ with $\sigma =\sqrt{5}/\kappa_{\rm m}$. After a storage time $t_{\rm storage}=t_{2}-t_{1}, t_{2}=5t_{\rm m}$, the second Gaussian allows us to transfer the quantum state back to the optical mode of the second cavity. Here we use $G_{0}/\kappa_{\rm m}=0.32$, $t_{\rm f}=100/\kappa_{\rm m}$ and $t_{\rm m}=t_{\rm f}/8$.}\label{fig5}
\end{figure}

The process of the quantum memory can still be described by Eqs. \eqref{l11}-\eqref{l13} with a pulsed optomechanical coupling of the form
\begin{align}\label{opt}
G(t)=G_{0}\left[e^{-(t-t_{1})^2/2\sigma^2}+e^{-(t-t_{2})^2/2\sigma^2}\right],
\end{align}
where $G_{0}$ is the maximum optomechanical coupling; $\sigma$ is the width of the Gaussian pulses and $t_{i}$ are the time at which the optomechanical coupling is maximum and the quantum state transfer from one mode to the other occurs. The first Gaussian pulse allows the transfer of the quantum state from the optical mode of the second cavity to the mechanical mode, while the second pulse allows the transfer of the quantum state back to the optical mode in the second cavity after a storage time $t_{\rm storage}=t_{2}-t_{1}$. The pulse sequence is shown in Fig. \ref{fig5} (a).

We numerically simulated the quantum memory using the pulsed optomechanical coupling \eqref{opt} and the time-varying damping rates, \eqref{k1} and \eqref{k2}. We see from Fig. \ref{fig5} (b) that the quantum state initially encoded onto the mode of the cavity 1 is transferred to the mode of cavity 2 with $99.4\%$ efficiency. Then, the quantum state is transferred to the mechanical oscillator by applying the first pulse of the coupling $G(t)$ [see Fig. \ref{fig5} (a)]. The efficiency of this transfer close to unity and decreases with increasing the storage time due to very small mechanical decoherence, $\gamma/\kappa_{\rm m}=6.5\times 10^{-4}$ ($Q=\omega_{\rm M}/\gamma\approx 6,700$). For $G_{0}=0.32 \kappa_{\rm m}$ the efficiency of the quantum memory after a storage time $t_{\rm storage}=5t_{\rm m}=5 t_{\rm f}/8=5\times 100/8\kappa_{\rm m}= 62.5/\kappa_{\rm m}$ ($46~\mu s$ for the experimental damping rate $\kappa_{\rm m}/2\pi= 215~\text{kHz}$) is $96\%$. To retrieve the quantum state from the mechanical oscillator, we apply the second pulse of the coupling. During this last procedure the quantum state is transferred to the mode of cavity 2, which will be released and measured via homodyne detection. The efficiency of quantum memory for a procedure time $t_{\rm f}=100/\kappa_{\rm m}$ is $96\%$. This efficiency exponentially decreases with increasing the procedure time as $\sim\exp(-\gamma t_{\rm f})$; for example, a $50\%$ increase in the procedure time leads to an efficiency of $94\%$. Thus the effect of the mechanical decoherence is not that significant for reasonably long storage time. This efficiency can substantially be improved by designing high-Q mechanical oscillator. For microwave resonators with high-Q mechanical oscillator, $Q=360,000$ and dissipation rates $\gamma_{\mu}/2\pi=30~\text{Hz}$ and $\kappa_{\mu}/2\pi=170~\text{kHz}$ \cite{Teu11}, the quantum memory efficiency improves to $99\%$ for a procedure time $t_{\rm f}=100/\kappa_{\mu}$.

Even though we used classical field equations to analyze the state transfer, it is sufficient to characterize the quantum case. For example, if the initial state is a superposition of Fock state (a qubit), $|\psi\rangle_{\rm in}=\alpha_{q}|0\rangle+\beta_{q}|1\rangle$ and the system is coupled to vacuum environment, the quantum process fidelity is related to the efficiency by $F_{\chi}=(1+\eta+2\sqrt{\eta}\cos \varphi)^2/4$, where $\varphi$ is the phase acquired by a photon carrying state and can be corrected experimentally \cite{Set14}. It is then easy to see that the quantum state transfer can be characterized by a single parameter $\eta$. For this simple case, the probability amplitudes of the wave function satisfy the same equations as the normalized classical field equations.
Arbitrary quantum state transfer (more than single excitation subspace) can be described by using the language of quantum theory of beam splitter \cite{Set14}. However, for thermal environment the classical equations are not sufficient to analyze the quantum state transfer. To properly study the effect of environmental thermal phonons, one needs to solve the master equation of the system, including dissipations. It was shown that although the thermal phonon bath reduces the overall state transfer fidelity, its effect can be compensated by improving the mechanical Q-factor \cite{Mcg13}.

\section{conclusion}
We analyzed transfer of a quantum state between an optical cavity and a mechanical oscillator coupled to a distant cavity via a transmission channel and vise versa. By employing time-varying cavity damping rates, it possible to achieve a state transfer between the two remote systems with efficiency very close to unity. We also proposed a quantum memory device using the mechanical oscillator. We showed that using experimental parameters and moderate mechanical Q-factor, the efficiency of the quantum memory can reach above $96\%$ with a storage time $t_{\rm storage}= 62.5/\kappa_{\rm m}$  with $\kappa_{\rm m}$ being the maximum damping rates of the cavities. Although the mechanical decoherence slightly decreases the efficiency of the quantum memory, its effect can be suppressed by designing high-Q mechanical oscillator. Given the advancement of the superconducting technology and the realization of high-Q mechanical oscillators \cite{Pal13,Teu11}, the proposed system can be realized using superconducting microwave resonators connected by a transmission line.

\begin{acknowledgements}
E.A.S acknowledges financial support from the
Office of the Director of National Intelligence (ODNI), Intelligence
Advanced Research Projects Activity (IARPA), through the Army
Research Office Grant No. W911NF-10-1-0334. All statements of fact,
opinion or conclusions contained herein are those of the authors and
should not be construed as representing the official views or
policies of IARPA, the ODNI, or the U.S. Government. He also
acknowledge support from the ARO MURI Grant No. W911NF-11-1-0268.
\end{acknowledgements}

 \end{document}